**Symmetry and Nonstoichiometry as Possible Origin of Ferromagnetism in Nanoscale oxides**

Takashi Uchino
*Department of Chemistry, Graduate School of Science, Kobe University, Nada, Kobe 657-8501, Japan*

Toshinobu Yoko
*Institute for Chemical Research, Kyoto University, Uji, Kyogo 611-0011, Japan*

We show through density functional theory calculations that extended magnetic states can inherently occur in oxides as the size of the crystals is reduced down to the nanometer scale even when they do not explicitly include intrinsic defects. This is because in nanoscale systems crystallographically perfect crystallites paradoxically result in nonstoichiometric compositions owing to the finite number of constituting atoms. In these structurally perfect but stoichiometrically imperfect nanocrystallites, the spin-triplet state is found to be more stable than the spin-singlet state, giving rise to an extended spin distribution that expands over the entire crystal. According to this picture, long-range magnetic order arises from the combined effect of crystal symmetry and nonstoichiometry that can coexist exclusively in nanoscale systems. The idea can also give reasonable explanations for the unprecedented ferromagnetic features observed commonly in nanoscale oxides, including ubiquity, anisotropy, and diluteness.

Magnetism in solids was thought to be well-understood in terms of localized magnetic moments *m* and a Heisenberg exchange integral *J*. In oxides, the principal interaction stems from superexchange, in which the effective exchange coupling of magnetic atom occurs via an intervening non-magnetic oxygen atom. The superexchange interaction is short-ranged and generally favors antiferromagnetic coupling. However, this *m-J* paradigm has recently been challenged by the finding of dilute magnetic semiconductors in which a few percent of the nonmagnetic cations are replaced by $3d$ transition-metal ions [1–3]. Subsequent investigations have further revealed that ferromagnetism can be found in closed-shell oxides doped with nonmagnetic elements or even in non-doped oxides [4–7]. It should also be worth mentioning that the ferromagnetism is found in a wide range of oxides, including MgO [8], $Al_2O_3$ [6], ZnO [6] $HfO_2$ [4,5], and $TiO_2$ [5], with different crystallographic structures and compositions. However, there is one common tendency among the materials exhibiting the intriguing ferromagnetism; that is, they are mostly in the form of thin films or nanoparticles [7]. This suggests that in nanoscale systems there exists an unrevealed underlying mechanism for ferromagnetism that is different from the conventional *m-J* paradigm [2].

What is unusual about the nanoscale system? One immediate possibility would be a mechanism related to defects [9,10]. It is indeed true that some intrinsic defects, e.g., cation and oxygen vacancies, provide the paramagnetic ($S=1/2$) states or the spin triplet ($S=1$) states. However, these defect-related spin states are generally highly localized around the respective defects, hence requiring a sufficiently high concentration to allow possible ferromagnetic order via magnetic percolation. At present, there have been no credible mechanisms that can account for the observed long-range ferromagnetic interactions in terms of the defect-related magnetic states [2,7,11].

The other possibility, which, however, has not so far been seriously discussed and considered, is the effect of the finite number of atoms in nanocrystals. Consider an oxide with a composition of $M_xO_y$. This oxide requires $nx$ atoms for M (where $n$ is integer) and $ny$ atoms for O to satisfy the stoichiometry. For example, MgO and $Al_2O_3$ crystals must consist of $2n$ and $5n$ atoms, respectively, to retain the stoichiometric composition. It should be noted, however, that this condition will not always be satisfied in the case of nanocrystals since the total number of constituent atoms is finite; it becomes odd or even depending on the size. As for the cubic rock-salt crystal structure, for example, crystallites consisting of an $N \times N \times N$-atom block become stoichiometric when $N$ is even, but an odd number of $N$ leads to, rigorously speaking, nonstoichiometric. In the latter case, the number of metal atom is always larger (or smaller) than that of oxygen atom by 1. The thus induced stoichiometric variation is very small and is virtually negligible in the case of the bulk where the number of atoms can be regarded as infinite. However, this might not be the case for the nanocrystals made up of atoms on the order of a few hundred to a few thousand. Only one atom difference between metal and oxygen atoms could induce a noticeable nonstoichiometry effect on the resulting electronic structure.

To explore a possible nanometer-sized nonstoichiometry effect we carry out a series of density functional theory (DFT) calculations using clusters of atoms modeling the cubic MgO nanocrystallites with different compositions. Previously, a number of theoretical calculations on MgO nanoclusters have been performed [12–15] since MgO is often considered as a prototype of ionic oxides. It has been demonstrated that the cubic rock-salt model can be applied to MgO nanoclusters when going beyond the size range of ~50 atoms [15]. However, most of the MgO clusters investigated previously are stoichiometric ones with an even number of atoms [12–15], and, to our knowledge, the electronic structure of the "cubic nonstoichiometric" clusters, i.e., cubic clusters consisting of an odd number of atoms, has not been theoretically examined.

All the DFT calculations in this work were carried out using the gradient corrected Becke's three parameters hybrid exchange functional [16] in combination with the correlation functional of Lee, Yang, and Parr (B3LYP) [17] with the standard 6-31G(d) basis set. It has previously been shown that such a hybrid DFT functional is useful to correct the self-interaction problem [18] which often leads to misleading conclusions with regards to hole localization and the resulting magnetic characteristics of the system [11,19]. We employed ideal cubic clusters of two different sizes (see Fig. 1), a stoichiometric cluster with a (4×4×4)-atom block and a nonstoichiometric cluster with a (5×5×5)-atom block. As for the 5×5×5 cluster, we assumed two nonstoichiometric compositions; that is, the $Mg_{62}O_{63}$ (model I) and $Mg_{63}O_{62}$ (model II) clusters, in which the central atom is O and Mg, respectively. Thus, models I and II correspond respectively to the Mg-deficient and O-deficient compositions. All the Mg–O interatomic distances in these model clusters were fixed to the experimental value (2.1056 Å) of the bulk MgO crystal [20]. Under this fixed structural condition, we carried out DFT calculations of the respective clusters for both the spin singlet ($S$=0) and triplet ($S$=1) states with the GAUSSIAN-09 [21] code.

As for the stoichiometric 4×4×4 cluster, we found that the non-magnetic $S$=0 ground state was correctly predicted, in agreement with the general consensus that MgO is a diamagnetic oxide. The energy separation between the lowest singlet ($S$=0) and triplet ($S$=1) states is 2.4 eV, which is certainly too large to anticipate any magnetism from the stoichiometric cluster.

We turn to the results of the nonstoichiometric 5×5×5 clusters (models I and II). In contrast to the case of the stoichiometric cluster, the magnetic ($S$=1) state was predicted to be lower than the non-magnetic ($S$=0) state for both models I and II. The energy separations between the $S$=1 and $S$=0 states are 0.487 and 0.175 eV for models I and II, respectively. It is hence probable that cubic but nonstoichiometric MgO clusters generally have the magnetic ($S$=1) ground state, raising the possibility that the built-in nonstoichiometry is responsible for the formation of magnetic moments in nanoscale oxides.

We next analyze the total spin density distribution, which is defined as the local density difference between the spin-up and spin-down states, in the magnetic spin state ($S$=1). First, we investigate the result of model I [see Fig. 2(a) and (b)]. Although the most of the spin density resides on the corner O atoms, the spin density distribution also spreads into the inner O atoms in the cluster, showing a symmetrical and extended nature of the spin density distribution because probably of the assumed cubic symmetry of the cluster. The preferential location of the spin density on O atoms can be interpreted in terms of the composition of model I ($Mg_{62}O_{63}$), namely, the Mg deficient composition. In an ionic picture, with $Mg^{2+}$ and $O^{2-}$ as closed-shell ions, deficiency in one Mg atom in the neutrally charged system would have two holes, which are expected to be localized preferentially at the fully filled $2p$ orbitals of oxygen [22]. The expectation is in harmony with the molecular orbital energy-level diagrams shown in Fig. 3(a). In the $S=0$ spin state, both the highest occupied molecular orbital (HOMO) and lowest unoccupied molecular orbital (LUMO) have O $2p$ characters. This is also true for the spin $S=1$ state. In the spin $S=1$ state, however, the spin-up HOMO level is lower than the spin-down HOMO level, and the spin-down LUMO level, which is a doubly degenerate state, is higher than the spin-down HOMO level only by 0.24 eV. The thus obtained electronic structures hence shows typical spin polarization features achieved by hole doping [23]. Thus, in model I, two holes located in the doubly degenerate spin-down LUMO are responsible for the extended nature of spin density. This situation is quite different from that of O-$2p$ holes in a conventional cation vacancy in oxides, where the two holes reside completely on two adjacent oxygen sites via polaronic distortion [22,24].

Such an extended nature of spin density can also be seen in model II with the O-deficient composition [see Fig. 2(c) and (d)]. As compared with the case of model I, however, the spin density in model II is much more evenly distributed over the entire cluster. In model II, the spin density is located mainly on the corner Mg atoms indeed, but the distribution is very broad, extending to the first-neighbor O atoms and even to the higher order nearest neighbors. This is because in model II both the HOMO and the LUMO levels consist not only of Mg 3$s$ and 3$p$ states but also of O 3$s$ and 3$p$ states, all of which are in principle empty levels in their closed-shell ions, $Mg^{2+}$ and $O^{2-}$. Thus, in model II, the spin polarization is achieved by electron doping; the two electrons occupying the two highest spin-up HOMO levels contribute to the extended spin state [see Fig. 3(b)]. Note also that in model II the two highest spin-up HUMO levels are energetically well separated by the third highest HOMO level by 3.24 eV, yielding the spin polarized state analogous to the one due to donor impurities [23]. These features contrast sharply with those of a normal oxygen vacancy where the singlet $S$=0 state is always the lowest and no magnetic moments are to be expected [7].

All the model clusters employed above do not explicitly contain defect centers, thus keeping the ideal cubic symmetry. In the real nanoscale systems, however, it is most likely that some sorts of intrinsic defects are introduced especially at their surface. As for the MgO surface, it has been well recognized that the most prototypical and stable defect is the Schottky pair, namely, the neutral divacancy consisting of a magnesium vacancy and an oxygen vacancy [25,26]. It is hence interesting to evaluate the effect of the introduction of divacancy on the inherently nonstoichiometric 5×5×5 cluster. For that purpose, we removed a pair of Mg and O atoms from one of the surfaces of model II to form the $Mg_{62}O_{61}$ cluster [see Fig. 4(a)] and carried out single point DFT calculations for the singlet and triplet states. The DFT calculations also predicted that the ground state of model III is the magnetic ($S$=1) state. The energy separation between the magnetic and non-magnetic ($S$=0) states is 0.176 eV, which is very similar to the one obtained for model II (0.175 eV) mentioned earlier. Furthermore, the extended nature of the spin density is still preserved in model III although the spin density tends to be localized within the divacancy site [see Fig. 4(b)]. This implies that the basic conclusions obtained from the ideal cubic clusters can also be applied to the clusters containing intrinsic defects as long as they retain a high degree of structural symmetry and nonstoichiometry.

Thus, the mechanism of magnetism predicted in the cubic nonstoichiometric clusters is fundamentally different from the conventional one based on the *m-J* paradigm. According to the *m-J* paradigm, ferromagnetic order is formed on a local-by-local basis via an exchange integral, or in a bottom-up manner. In nanoscale systems, however, the extended magnetic states are likely to be realized through a top-down strategy. That is, the extended nature is inherently built-in in the symmetry of nanocrystals, and the spin polarization is derived from the destined compositional deficiency. In that sense, ferromagnetism can be found in a variety of nanoscale oxides irrespective of the chemical composition, probably accounting for the reported ubiquitous feature of ferromagnetism [4–8]. Also, the top-down model allows us to assume that the direction of magnetic moment is determined by the symmetry and/or the shape of each crystallite. Thus, the anisotropy of the magnetization observed often in polycrystalline thin films [4] can be interpreted in terms of crystallographic anisotropy and/or the presence of a preferential growth direction in each crystallite. The present model may also give a reasonable account why dilute magnetic oxides exhibit ferromagnetism only at doping concentrations well below the percolation threshold [2,27]. It is reasonable to expect that magnetization inherent to the nanocrystals will be enhanced by doping of atoms, either magnetic or non-magnetic, because doping will somehow promote nonstoichiometry. However, the larger the doping concentration the lower the symmetry of the host crystal. Thus, the heavy doping up to the percolation threshold is not favorable in view of the formation of the proposed symmetry-driven ferromagnetism.

In conclusion, we put forward a model of ferromagnetism based on "symmetry" and "nonstoichiometry," which can in principle be applied to every nanocrystalline system. The model does not assume any long-range connectivity of local magnetic moments due to specific dopants and/or defects; rather, it predicts that the extended magnetic states are inherent characteristics of nanocrystals with a high structural symmetry and a stoichiometric deficiency. We believe that the proposed model will shed new light on the origin of ferromagnetism not only in non-doped closed shell oxides but also in dilute magnetic semiconductors.

We thank the Supercomputer System, Institute for Chemical Research, Kyoto University, for providing the computer time to use the SGI Altix 4700 supercomputer.

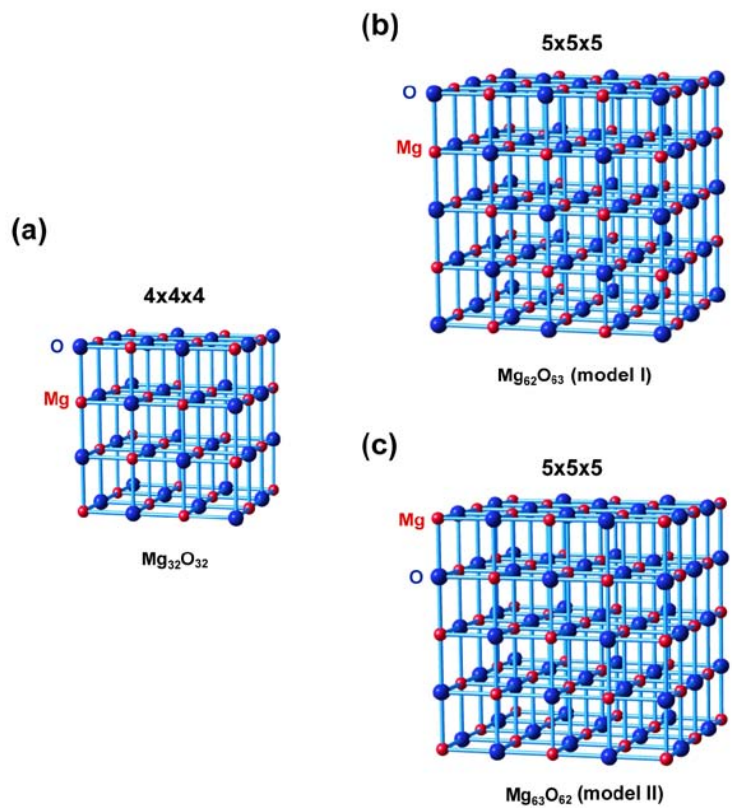

**Figure 1.** Cubic cluster models of MgO crystal. All the Mg–O interatomic distances are fixed to the experimental value (2.1056 Å) of the bulk MgO crystal. (a) A (4×4×4)-atom block of stoichiometric composition $Mg_{32}O_{32}$. (b) A (5×5×5)-atom block of nonstoichiometric composition $Mg_{62}O_{63}$ (model I) in which the central atom is oxygen. (c) A (5×5×5)-atom block of nonstoichiometric composition $Mg_{63}O_{62}$ (model II) in which the central atom is magnesium.

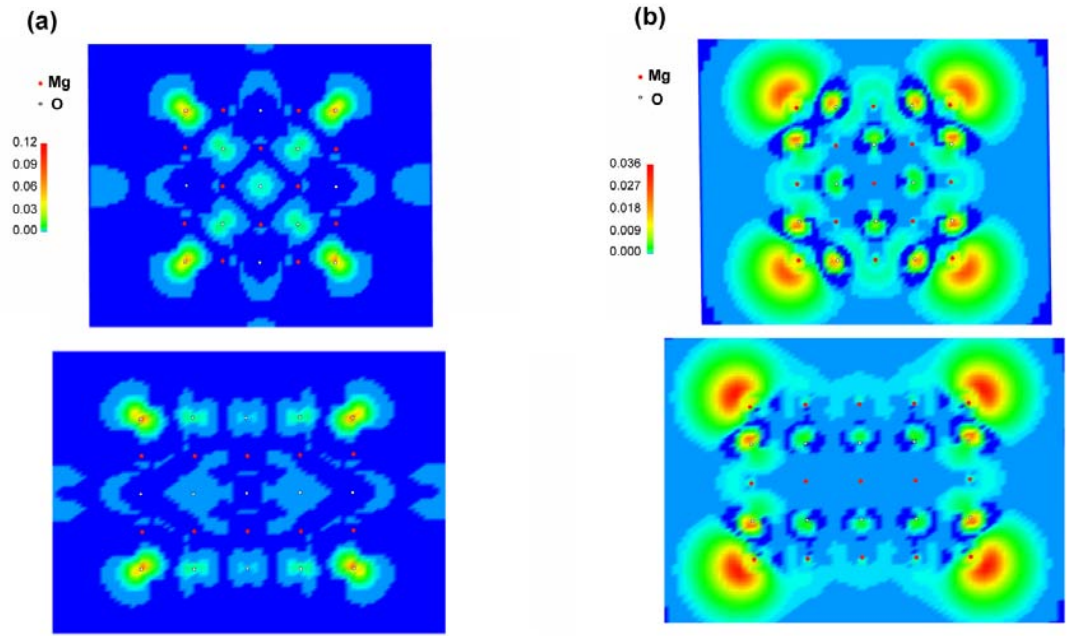

**Figure 2.** Spin density maps of (a) model I and (b) model II calculated for the magnetic ($S$=1) ground state. Upper and lower panels correspond to the results on the (100) and (110) planes, respectively.

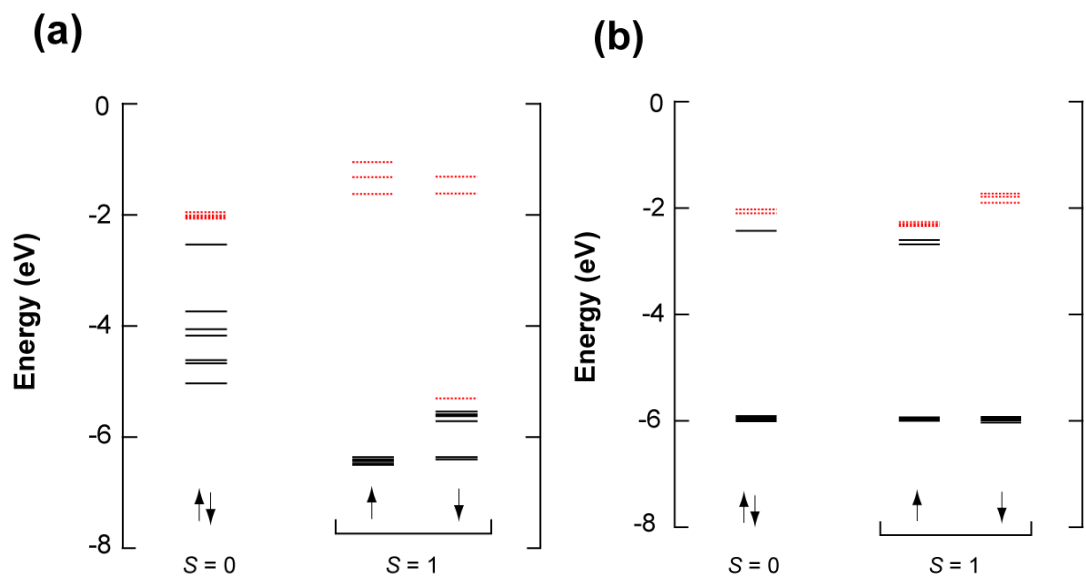

**Figure 3. (color online)** Molecular orbital energy-level diagram of the ten highest HOMOs (black solid lines) and the five lowest LUMOs (red dotted lines) obtained for a non-magnetic ($S$=0) state and a magnetic ($S$=1) ground state of (a) model I and (b) model II.

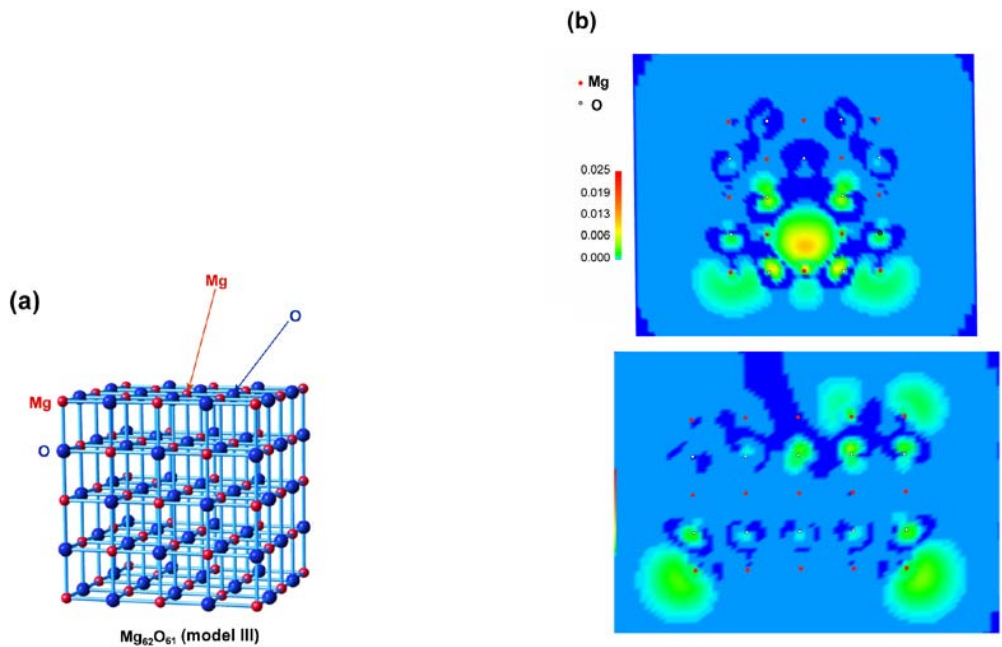

**Figure 4.** (a) A cluster model of defective cubic MgO (model III) based on a (5×5×5)-atom block in which the central atom is Mg. A pair of Mg and O atoms, which are located in the (001) plane as indicated by arrows, are removed from model II, yielding nonstoichiometric composition $Mg_{62}O_{61}$. (b) Spin density maps of model III calculated for the magnetic ($S$=1) ground state. Upper and lower panels correspond to the results on the (001) and (110) planes, respectively. In the (110) plane, the supposed divacancy pair is located in the upper right side.